\documentclass[sigconf]{acmart}
\usepackage{array}
\usepackage{multirow}
\usepackage{caption}

\AtBeginDocument{%
  }

\copyrightyear{2025}
\acmYear{2025}
\setcopyright{cc}
\setcctype{by}
\acmConference[CHI '25]{CHI Conference on Human Factors in Computing Systems}{April 26-May 1, 2025}{Yokohama, Japan}
\acmBooktitle{CHI Conference on Human Factors in Computing Systems (CHI '25), April 26-May 1, 2025, Yokohama, Japan}\acmDOI{10.1145/3706598.3713626}
\acmISBN{979-8-4007-1394-1/25/04}

\begin{document}

\title[Sustaining Human Agency, Attending to Its Cost]{Sustaining Human Agency, Attending to Its Cost: An Investigation into Generative AI Design for Non-Native Speakers' Language Use}

\author{Yimin Xiao}
\affiliation{%
  \department{College of Information}
  \institution{University of Maryland}
  \city{College Park}
  \state{Maryland}
  \country{USA}
}
\email{yxiao@umd.edu}

\author{Cartor Hancock}
\affiliation{%
  \department{UMIACS}
  \institution{University of Maryland}
  \city{College Park}
  \state{Maryland}
  \country{USA}
}
\email{cartorh@terpmail.umd.edu}

\author{Sweta Agrawal}
\affiliation{%
  \institution{Instituto de Telecomunicações}
  \city{Lisbon}
  \country{Portugal}
}
\email{swetaagrawal20@gmail.com}

\author{Nikita Mehandru}
\affiliation{%
  \department{School of Information}
  \institution{University of California, Berkeley}
  \city{Berkeley}
   \state{California}
  \country{USA}
}
\email{nmehandru@berkeley.edu}

\author{Niloufar Salehi}
\affiliation{%
  \department{School of Information}
  \institution{University of California, Berkeley}
  \city{Berkeley}
  \state{California}
  \country{USA}
}
\email{nsalehi@berkeley.edu}

\author{Marine Carpuat}
\affiliation{%
  \department{Department of Computer Science}
  \institution{University of Maryland}
  \city{College Park}
  \state{Maryland}
  \country{USA}
}
\email{marine@umd.edu}

\author{Ge Gao}
\affiliation{%
  \department{College of Information}
  \institution{University of Maryland}
  \city{College Park}
  \state{Maryland}
  \country{USA}
}
\email{gegao@umd.edu}

\renewcommand{\shortauthors}{Yimin Xiao et al.}



\begin{abstract}
AI systems and tools today can generate human-like expressions on behalf of people. It raises the crucial question about how to sustain human agency in AI-mediated communication. We investigated this question in the context of machine translation (MT) assisted conversations. Our participants included 45 dyads. Each dyad consisted of one new immigrant in the United States, who leveraged MT for English information seeking as a non-native speaker, and one local native speaker, who acted as the information provider. Non-native speakers could influence the English production of their message in one of three ways: labeling the quality of MT outputs, regular post-editing without additional hints, or augmented post-editing with LLM-generated hints. Our data revealed a greater exercise of non-native speakers’ agency under the two post-editing conditions. This benefit, however, came at a significant cost to the dyadic-level communication performance. We derived insights for MT and other generative AI design from our findings. 
\end{abstract}



\maketitle

\section{INTRODUCTION}
As it is increasingly feasible for AI-powered systems and tools to generate natural-sounding sentences and expressions, the question of how much humans can, and need to, exercise their own agency in communication deserves more attention than ever. Recent research in the HCI community has contributed valuable insights to this discussion. Specifically, existing work has cautioned against the mindless adoption of AI-generated materials (e.g., \cite{reza2024ABScribe, li2024value, gero2023social, kim2024diarymate, xiao2024displacedcontributionsuncoveringhidden}). It has also demonstrated cases where people strike a good balance between “benefiting from AI usage” and “sustaining their agency during AI usage” by closely monitoring and evaluating AI outputs before using them (e.g., \cite{fu2023comparing, fu2024text, lee2022coauthor, Rezwana2023Designing}). 

In the current research, we pay attention to machine translation (MT), the AI system that promises to mitigate language gaps in conversations. Prior work with non-native speakers of a society’s dominant language (e.g., English) suggests that these individuals, often being new immigrants to a country, sometimes rely on MT to perform conversations with local native speakers \cite{lieblingUnmetNeedsOpportunities2020a,adjagbodjouEnvisioningSupportCenteredTechnologies2024, gao2017kaleidoscope}. However, the use of MT is often accompanied by a loss of human agency: non-native speakers can effectively formulate thoughts and put them into words in their native language, but the representation of their meanings in English is entirely controlled by MT. The human-MT interface today often provides non-native speakers with limited opportunities to influence MT outputs \cite{wang2013machine, robertson2022understanding, yamashita2006effects,chenLinguaFrancaSystemFacilitated2018, gaoHowBeliefsPresence2014}. 

Inspired by the agency-resource connection discussed in Bandura’s social cognitive theory \cite{bandura2006toward, bandura1999}, we explore whether non-native speakers can leverage their own language resources in English - that is, the English vocabulary they have acquired to comprehend and produce English content - to influence MT outputs and, hence, better sustain their agency in MT-mediated conversations with native speakers. More specifically, we suspect that, given the unique composition of non-native speakers’ English resources, future human-MT interfaces could be designed to facilitate the co-production of English messages by a non-native speaker and the MT in more than one way. We conducted an experimental study to compare the effects of three possible designs of this human-MT interface: a) the labeling interface, where non-native speakers can leverage a subset of their English resources (e.g., passive vocabulary) to mark their perceived quality of MT outputs for a conversation partner to view, b) the regular post-editing interface, where non-native speakers can leverage the full set of their English resources (e.g., passive and active vocabulary) to revise the MT outputs before relaying them to a communication partner, and c) the augmented post-editing interface, where non-native speakers can access large language model (LLM) generated hints as an additional resource to facilitate their revision of MT outputs. The comparison across these interface conditions allowed us to explore possible ways to sustain the non-native speakers’ agency while remaining cautious about the associated costs. 

The rest of this paper details the rationale, method, data analysis, and results of our research. In particular, our research indicated that, although all three human-MT interface designs could help sustain the non-native speakers' agency at different extents, the rise of their agency under the two post-editing interface conditions came at a greater cost to people’s communication process and outcome. Compared to those assigned to the labeling interface condition, dyads under the two post-editing interface conditions exchanged more information in their conversations; however, these conversations were characterized by lower depth and resulted in less alignment established between interlocutors. These results shed light on future human-MT interaction design that can facilitate non-native speakers in leveraging their (constrained) English resources to sustain agency, as well as managing the trade-offs between agency and other essential aspects of communication.

\section{RELATED WORK}
In this section, we review previous literature that informs the hypotheses (Hs) and research questions (RQs) guiding our work. We begin with Bandura’s theory of human agency, which constitutes an effective device for understanding AI-mediated communication (Section 2.1). Next, we focus on the use case of MT, considering the “agency loss” experienced by those who often rely on translation to navigate information-seeking conversations in their non-native language (Section 2.2). Building upon this literature, we investigate ways to sustain non-native speakers’ agency in MT-assisted communication. Our point of departure is a comparison between three possible conditions of human-MT interfaces. They all have the potential to protect non-native speakers’ agency in a conversational setting, but likely to varying extents and with different associated costs (Section 2.3). 

\subsection{Bandura's Theory and Its Implications for Agency in AI-Mediated Communication}

Agency considers a person’s capability to influence the achievement of desired outcomes \cite{bandura2006toward}. Among the bulk of social science theories tackling this concept (e.g., \cite{emirbayer1998agency,ahearn2001language, schwartzUniversalPsychologicalStructure1987, ryffPonceLeonLife1989}), arguably the most influential one was proposed by Bandura \cite{bandura1999, bandura2006toward}. His social cognitive theory emphasizes that people are not merely onlookers of what happens to them but active contributors to the world they live in \cite{bandura2001social}. Rather than passively reacting to the environment, individuals demonstrate personal agency by evaluating opportunities and constraints and planning actions accordingly \cite{banduraSocialCognitiveTheory1991, lockeTheoryGoalSetting1990}. 

At the heart of Bandura’s conceptualization lies the connection between personal agency and the resources an individual can utilize – such as knowledge and materials within their own access – to take purposeful actions toward their goal \cite{bandura1999, bandura2006toward, banduraSelfEfficacyMechanismHuman1982}. A lack of such resources forces people to be “wholly determined” by situational influences \cite{ bandura2006toward}. Alternatively, the person may turn to other individuals or groups who possess the resources to act on their behalf, which places the self in “a vulnerable position resting on the competencies and favors of others \cite{banduraSelfEfficacyMechanismHuman1982, banduraPsychosocialImpactMechanisms2003}”. 

The agency-resource connection provides an inspiring theoretical lens for studying AI-mediated communication. In particular, a rich body of recent HCI research has cautioned that the rise of generative AI today may challenge the preservation of users’ agency (e.g., \cite{fu2023comparing, kim2024diarymate, kobiella2024if, gero2023social, li2024value}). Much of this challenge, according to Bandura’s theory, stems from AI’s increasing ability to generate human-like resources, such as natural-sounding sentences and expressions, that are essential for task completion. For instance, Fu et al. conducted an experiment with 120 participants who completed email writing with the help of AI. Participants in this study reported a greater loss of personal agency when the AI drafted the entire email for them, rather than providing specific suggestions upon the user’s requests and for their review \cite{fu2023comparing}. Kim et al. interviewed 24 individuals who engaged in AI-assisted personal journaling over a 10-day period. While all interviewees expressed enthusiasm for this practice, they voiced concerns that the blunt adoption of AI-generated language would prevent people from documenting genuine thoughts and feelings in their journals \cite{kim2024diarymate}. Kobiella et al. conducted a longitudinal study with knowledge workers to understand their use of AI for content creation and information discovery. People reported a diminished sense of control when AI’s contribution to the task overshadowed the person’s own effort \cite{kobiella2024if}. Gero et al. and Li et al. examined the agentic dynamics between human and AI in various cases of essay writing. Participants in both studies perceived their language use to be less authentic as they used AI assistance in less reflective manners \cite{gero2023social, li2024value}. 

Together, the above literature underscores the necessity for humans to utilize their own resources toward task completion. In AI-mediated communication, good practices of technology design should allow or even encourage people to monitor, evaluate, and influence AI-generated resources, rather than have the AI act on their behalf.   

\subsection{Non-Native Speakers' Agency in the Use Case of Machine Translation}
Our current research explores ways to sustain human agency in the use case of machine translation (MT), the AI system developed to bridge language gaps in communication. This choice situates our empirical discussion about human agency in a unique setting. Specifically, prior HCI research on MT-assisted conversations often involved individuals who aim to navigate information exchanges in their non-native language, such as immigrants with imperfect fluency in a society’s dominant language but need to communicate their information needs with local native speakers (e.g., \cite{lieblingUnmetNeedsOpportunities2020a,adjagbodjouEnvisioningSupportCenteredTechnologies2024, duUnderstandingInformationJourneys2023, adkinsInformationBehaviorICT2020a,beretta2018immigrants,baron2014crossing, gao2022taking}). Numerous studies have demonstrated the potential of MT in assisting non-native speakers’ message production in the target language (e.g., \cite{lieblingUnmetNeedsOpportunities2020a, adjagbodjouEnvisioningSupportCenteredTechnologies2024, adkinsInformationBehaviorICT2020a,lingelInformationTacticsImmigrants2011,baron2014crossing} ); that said, the default setup of human-MT interface always compels non-native speakers to accept the system’s translation as-is, even at times when people are not fully satisfied with the translation. Their personal agency in MT-assisted conversations is noticeably limited. 

The agency-resource connection, as conceptualized by Bandura, points to a direction for better preserving non-native speakers’ agency: non-native speakers should be allowed to influence MT outputs with the language resources they own; accordingly, the human-MT interface could be designed to enable this influence. While existing HCI literature has rarely discussed such interface design for the agency of non-native speakers, we notice relevant examples in the field of translation studies (e.g., \cite{cadwell2016human, guerberof2013professional, koponen2016machine}). In that context, professional translators employ MT as a productivity accelerator. They let MT take the first pass at processing the source text, then review the MT’s output and post-edit it by leveraging their own language knowledge. As O’Brien’s recent article notes, the post-editing process enables professional translators to speed up their workflow while maintaining “their intended production \cite{obrienHowDealErrors2022}.” To further support professional translators in performing post-editing, several HCI projects have developed and tested tools tailored to this user group’s needs. Recent examples include Green and colleagues' work on predictive translation memory \cite{green2014predictive,greenEfficacyHumanPostediting2013a}, Coppers et al.'s development of Intellingo \cite{coppers2018intellingo}, and Herbig and coauthors' system for multimodal translation editing \cite{herbig2019multi,herbigMMPEMultiModalInterface2020}.

So, is post-editing also an effective way to sustain non-native speakers’ agency in the use of MT? Our literature review suggests that the answer may not be straightforward. Unlike professional translators who are fluent bilinguals in both the source and the target languages, non-native speakers possess constrained resources in the target language. This constraint has been extensively evidenced in communication and psycholinguistics research on the “recognition and production gap (e.g., \cite{gass1983development})”. On the one hand, being a non-native speaker often means the person has passive vocabulary to sense expression problems in the target language, allowing for their self-initiated attempts to repair those problems \cite{hosoda2000other, day1984corrective, chun1982errors, fernandez2014native}. On the other hand, this person may or may not own sufficient active vocabulary to complete the repair \cite{chun1982errors, fernandez2014native, kurhila2001correction, wong2000delayed, hautasaari2014maybe}. Previous studies have found that, in order to improve the quality of writing in the target language, non-native-speaking writers sometimes consult model texts to enhance their awareness of a greater variety of expressions (e.g., \cite{gass1983development, ferris2013written}). Other research has also documented instances where non-native speaking employees seek suggestions from native speaking colleagues to refine their oral presentations in the working language (e.g., \cite{neeley2013language, aichhornJustDontFeel2017}). 

In summary, prior literature implies that non-native speakers have the language resources to influence MT outputs and, thereby, better maintain their agency in MT-assisted conversations. However, the composition of these resources differs from that of fluent bilinguals. Non-native speakers are generally more effective in utilizing a subset of their knowledge in the target language (e.g., passive vocabulary) for recognition, such as assessing a given expression as they encounter it. They tend to be less effective in utilizing their full range of knowledge in that language (e.g., passive and active vocabulary) for production, such as generating new expressions from scratch. These insights prompt us to explore alternative forms of human-MT interface design that enable non-native speakers to activate their language resources in a stepwise manner, allowing for varying levels of influence on MT outputs. 

\subsection{Interface Design for Non-Native Speakers' Influence on Machine Translation Outputs}
We considered three forms of human-MT interfaces to facilitate information exchange between non-native and native speakers. These interfaces enable non-native speakers to influence MT outputs by utilizing their language resources across distinct scopes. One type is a \textit{labeling} interface, where non-native speakers can use icon clicking to indicate their recognition of each MT output's quality without directly editing them. Another type is a \textit{regular post-editing} interface, similar to those used by professional translators who can edit translation outputs without additional hints. The third type is an \textit{augmented post-editing }interface, where non-native speakers can access additional hints, such as expressions recommended by an LLM, to scaffold their post-editing of MT outputs. 

By introducing these three interfaces, we do not claim to have exhausted all possible designs for enabling non-native speakers to exert influence while using MT. Rather, these interfaces represent different systemic infrastructures that allow a person to act with varying levels of their resources along a spectrum. Non-native speakers are expected to utilize a smaller scope of their language resources in the labeling condition (i.e., passive vocabulary for recognition tasks only), the full scope of their resources in the regular post-editing condition (i.e., passive and active vocabulary for both recognition and production), and additional resources in the augmented condition (i.e., passive and active vocabulary with additional hints to better activate these resources). As the first line of inquiry in our investigation, we ask: How will non-native speakers’ exercise of agency vary across these interfaces? Based on the agency-resource connection, we further hypothesize that non-native speakers would exercise a greater sense of personal agency in MT-assisted conversations when they could utilize a larger volume of resources to influence MT outputs:

\textbf{H1a.}  \textit{Non-native speakers will exercise a greater sense of agency in MT-assisted conversations when the task interface enables them to perform \textit{regular post-editing} of MT outputs, compared to \textit{labeling} their quality assessment of the outputs}.

\textbf{H1b.}\textit{ Non-native speakers will exercise a greater sense of agency in MT-assisted conversations when the task interface enables them to perform \textit{augmented post-editing} of MT outputs with additional hints, compared to \textit{regular post-editing}.}

Moreover, recent discussions on AI-mediated communication have cautioned that enhancing human agency may come with associated costs \cite{sundar2020rise, bennettHowDoesHCI2023}. In the use case of MT, existing HCI research has been paying close attention to people’s performance, in terms of their communication process and/or outcomes, and workload as two productivity-centered measures (e.g., \cite{zhang2022facilitating, yamashita2006effects, wang2013machine, gaoTwoBetterOne2015}). While we value non-native speakers’ agency in MT-mediated conversations, we believe it is equally important to be mindful of any potential trade-offs between this and the productivity aspects of communication. We ask the following questions in our current research context: 

\textbf{RQ1a.} \textit{How will the dyadic level communication process, such as the breadth and depth of information exchanges, differ when non-native speakers can influence MT outputs through \textit{labeling}, \textit{regular post-editing}, or \textit{augmented post-editing}?}

\textbf{RQ1b.} \textit{How will the dyadic level communication outcome, such as the alignment between information seekers and providers, differ when non-native speakers can influence MT outputs through \textit{labeling}, \textit{regular post-editing}, or \textit{augmented post-editing}?}

\textbf{RQ2. }\textit{How will the task workload experienced by communicants differ when non-native speakers can influence MT outputs through \textit{labeling}, \textit{regular post-editing}, or \textit{augmented post-editing}?}

\section{METHOD}
We designed a 3 $\times$ 2 between-subject experiment with 90 participants to investigate our hypotheses and RQs. This sample size was confirmed through a power analysis, and it aligned well with the practices of prior HCI studies involving dyadic or triadic participants performing communication tasks (e.g., \cite{duan2019increasing, schneiders2024understanding, park2023importance, rae2013body, yong2024change, caine2016local}). Half of our participants were Chinese new immigrants to the United States. They were all interested in housing selection tips and suggestions but lacked sufficient language proficiency to seek this
information in English. The remaining participants were local residents in the United States, who had extensive experience in selecting local housing. They spoke English as their sole native language. 

We paired participants into 45 dyads. Each dyad consisted of one non-native speaker and one native speaker of English. They performed text-based information-seeking conversations focused on a housing selection task that we developed. The text communication tool used in our experiment was embedded with an MT module, enabling non-native speakers to compose messages expressing their thoughts and questions in Mandarin before receiving an automated translation in English. The non-native speaker in each dyad was provided with one of the three human-MT interfaces: labeling interface, regular post-editing interface, or augmented post-editing interface. This experimental design allowed us to understand how the involvement of a non-native speaker’s language resources would affect their agency in MT-assisted conversation, as the level of resource involvement varied according to the human-MT interfaces people used. Below, we provide the full details of our method. 

\subsection{Participants}
We recruited all non-native speaking participants by distributing online and physical flyers within multiple Chinese new immigrant communities in the United States. Among the 45 individuals who signed up and participated in this research, 29 were self-identified females and 16 were males. Their average age was 27 (S.E. = .73). According to their self-reports on 7-point scales, these participants had a medium-level English proficiency (M = 4.56, S.E. = .14; 1 = limited proficiency, 7 = full proficiency). They held limited existing knowledge about housing selection in the United States (M = 2.18, S.E. = .16; 1 = no knowledge at all, 7 = extensive knowledge). They had some experience of communicating with local native speakers in daily work and life settings (M = 5.47, S.E. = .20; 1 = no experience at all, 7 = extensive experience). They also had some experience with language technology, such as translation and English text editing tools (M = 4.31, S.E. = .27; 1 = no experience at all, 7 = extensive experience). 

Native speakers in our sample were recruited from Prolific and other public social media platforms to ensure scheduling availability with non-native speakers. Among these 45 individuals, 31 were self-identified females and 14 were males. Their average age was 42.69 (S.E. = 1.79). They all spoke English as their native language and with full proficiency. According to their self-reports on a 7-point scale, these participants had sufficient knowledge about housing selection in the United States (M = 5.69, S.E. = .16; 1 = no knowledge at all, 7 = extensive knowledge). Their average experience of communicating with new immigrants to the country and with language technology was 3.71 (S.E. = .28) and 3.07 (S.E. = .25), respectively (1 = no experience at all, 7 = extensive experience). 

All participants were compensated at an hourly rate of \$20.

\subsection{Task Design}

\begin{figure*}[h]
    \centering
    \includegraphics[width=1\linewidth]{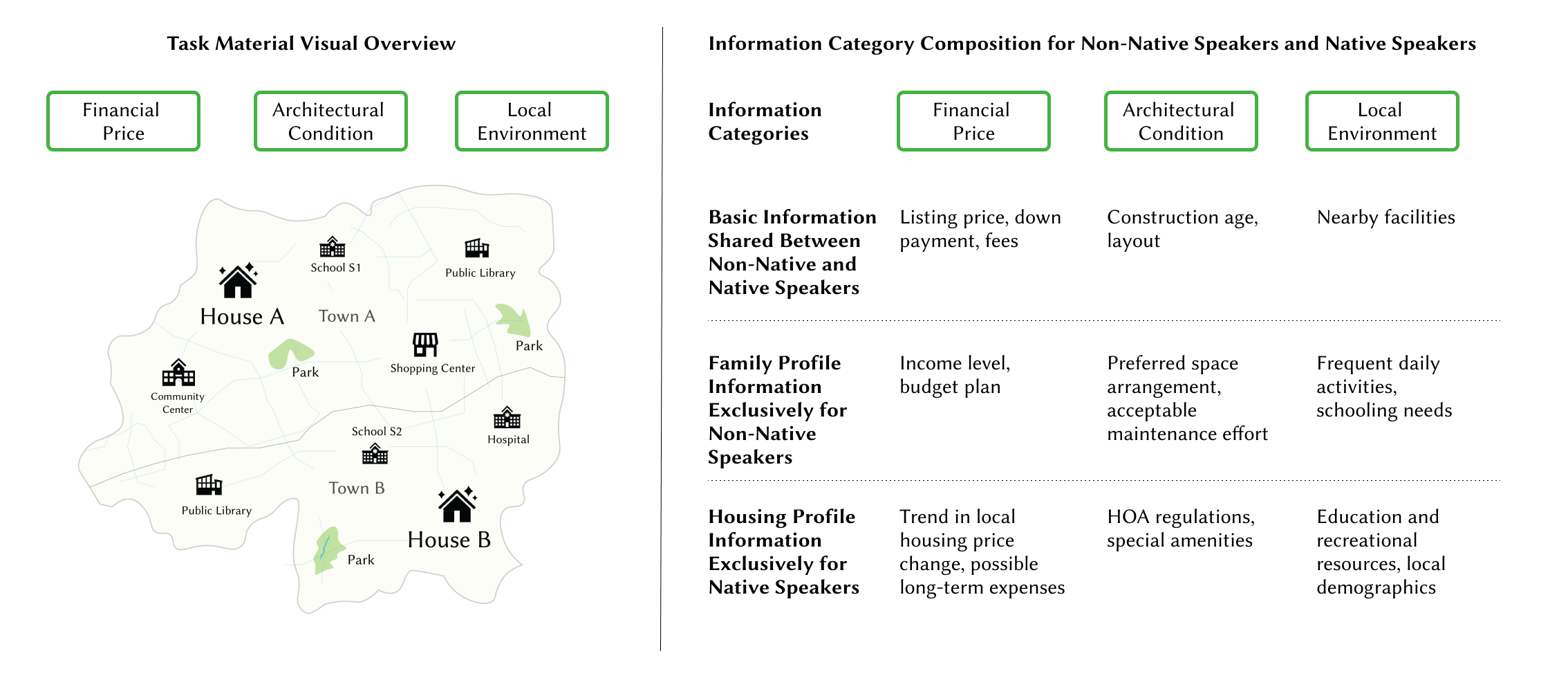}
    \caption{Composition of information in the task materials and its distribution between non-native speaking information seekers and native speaking information providers. As a participant clicks on each tab and housing options on the map view, the corresponding information will appear in a pop-up.}
    \label{fig:task_material}
    \Description{Figure 1 illustrates the information items we included in the task material and how these items were distributed between non-native speaking information seekers and native speaking information providers. On the left side is a sample page of our task material, where people can see tabs for three information categories: financial price, architectural condition, and local environment, and a map with the two housing options in their surrounding environment. Once participant clicks on each tab and housing options on the map view, the corresponding information will appear in a pop-up window. On the right side we list the sample information items under the three information categories. For each information category, we have basic items that were shared between non-native and native speakers (e.g., housing listing price and fees), family profile information exclusively for non-native speakers (e.g., family income level and budget plan), and housing profile information exclusively for native speakers (e.g., trend in local housing price change).}
\end{figure*}

\subsubsection{The housing selection scenario.} We developed an information seeking task within the housing selection scenario. New immigrants often face challenges in navigating essential information exchanges with local people in this context. This issue has been well-documented in prior research on and with immigrant populations \cite{beretta2018immigrants, caidi2010information, shoham2008immigrants}. To protect participants' privacy and ensure fair comparison across dyads in data analysis, we provided all participants with pseudo profiles that we prepared, rather than engaging in exchanges based on their personal information in real life. 

Our task design followed the established protocol of a “hidden profile task \cite{schulz2006group}.” Specifically, the task requires one non-native speaking participant to act the role of an information seeker, describing their housing needs and asking questions about the comparison between different housing options. In response, one native speaking participant, acting as the information provider, offered suggestions based on the housing information available to them and their knowledge from prior life experiences. Each party in this task was provided with a package of information that was hidden from the other party. This setup mimics the real-world information gap between communicants: non-native speaking participants had full access to descriptions about one pseudo information seeker’s family situation, whereas native speaking participants had full access to details about two housing options to discuss. It also allowed us to control for individual variations by situating all dyads within the same conversational context. Non-native speakers received their package of information in Mandarin, and native speakers received theirs in English. 

During the task conversation, participants shared information that their task partners would not have known otherwise. The ultimate goal of this information exchange was to bring both parties to a mutual understanding of the key considerations in making housing selections for the immigrant family’s needs. \autoref{fig:task_material} summarizes the types of information available to participants under the immigrant family’s profile and each housing option's profile, depending on their task role (i.e., information seeker or provider). 

\subsubsection{Task procedure} We paired participants in dyads based on their shared availability. Prior to the information-seeking task, each participant attended an online tutorial, getting themselves familiar with our task scenario and the task materials. We invited all non-native speaking participants to test how their message production would work with the human-MT interface designed for their task condition. Similarly, all native speaking participants were invited to preview how a task partner would use the MT and the translated message they would get. These steps helped ensure that our manipulation of the human-MT interface functioned as intended. 

During the information seeking task, dyads of participants completed their conversations about housing selection online. Each conversation session lasted 30 minutes and was performed entirely through text messages. The logs were recorded on our system server. Immediately after the conversation, the task interface directed participants to a short Qualtrics survey, which collected people’s numerical ratings of their conversation experiences across multiple aspects. Upon completing this survey, participants were provided with a Zoom link. They met the researcher for a brief session to receive instructions on compensation.

\subsection{System Setup} 

We developed an in-house communication tool for convenient recording and storage of all conversation logs. This tool features a front-end design that is identical to commonly available instant messaging applications. For the MT module integrated into this tool, we used an open-source Chinese-English translation model trained on the OPUS dataset, released by Helsinki-NLP \cite{TiedemannThottingal:EAMT2020}. The model achieves a BLEU score of 36.1 with a brevity penalty of 0.948 on the Tatoeba-test.zho.eng, indicating reasonable translation quality.

\begin{figure*}
    \centering
    \includegraphics[width=1\linewidth]{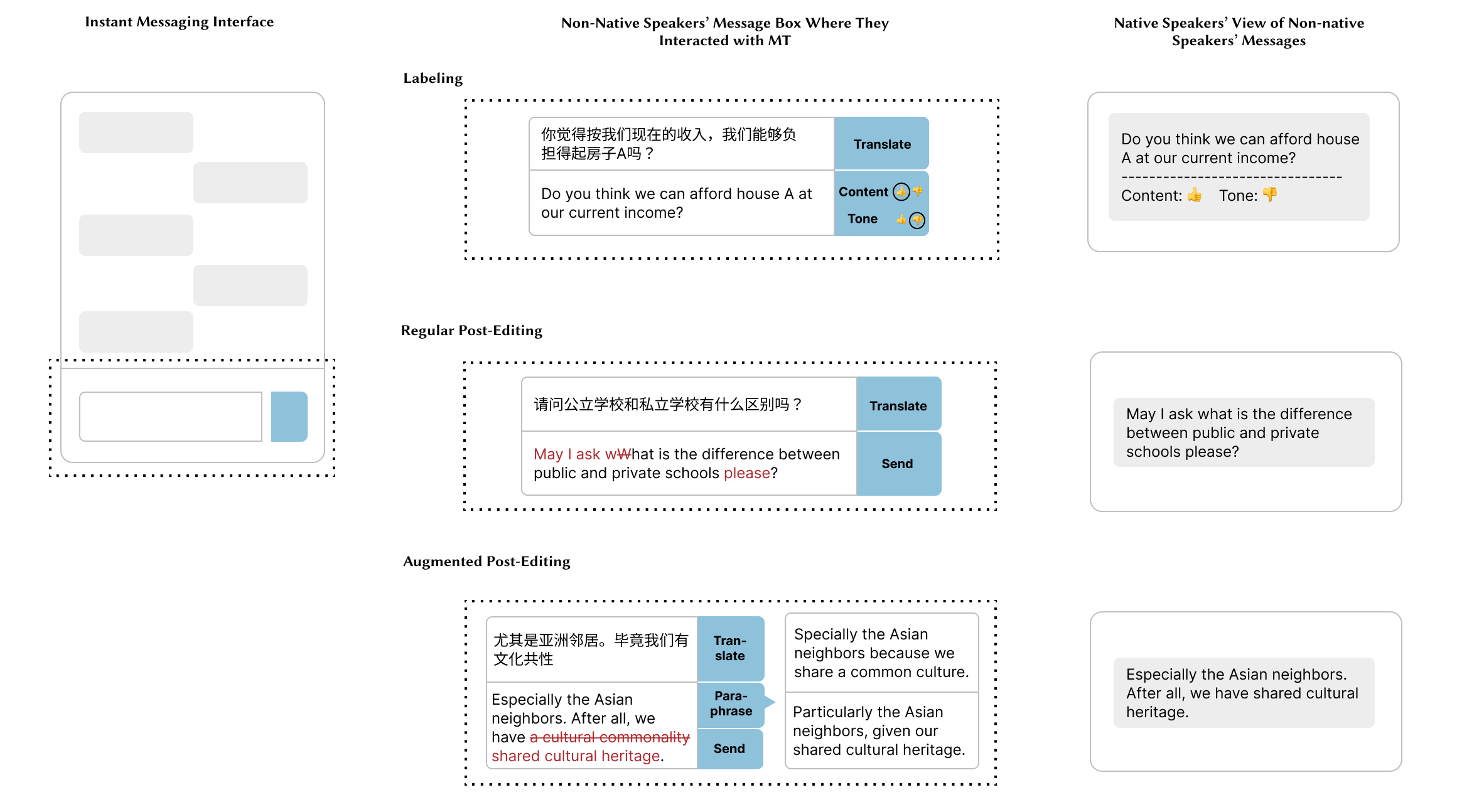}
    \caption{We manipulated how non-native speakers could influence their English production in the message box. In the labeling interface condition, non-native speakers could click on the thumbs-up and thumbs-down buttons to indicate their evaluation of the translation. Their native-speaking partners received both the English message and the evaluations. In the regular post-editing interface condition, participants could edit the original MT output in the English text box. In the augmented post-editing interface condition, participants could access two paraphrases of the initial MT output and edit it in the English text box. Their native speaking partners received the final version of English messages in the two post-editing conditions.}
    \label{fig:IM_interface}
    \Description{Figure 2 demonstrates our three human-MT interface conditions. We manipulated how non-native speakers could influence their English production in the message box. In the labeling interface condition, non-native speakers could click on the thumbs-up and thumbs-down buttons to indicate their evaluation of the translation. Their native speaking partners received both the English message and the evaluations. In the regular post-editing interface condition, participants could edit the original MT output in the English text box. In the augmented post-editing interface condition, participants could access two paraphrases of the initial MT output and edit it in the English text box. Their native speaking partners received the final version of English messages in the two post-editing conditions.}
\end{figure*}

\autoref{fig:IM_interface} illustrates the human-MT interface used by non-native speaking participants in each of the three task conditions, along with the corresponding view that native-speaking participants experienced when reading their task partners' messages. All participants viewed their side of the conversation history in a language setting that matched their own language profile: native speakers would view all messages in English as the sole language that makes sense to them, while non-native speakers would always see each message displayed in two languages (e.g., the translated version and the source version). This design choice aligned with recommendations from previous work on MT-assisted conversations (e.g., \cite{wang2013machine, zhang2022facilitating}), considering that non-native speakers often prefer to view a translated message alongside its source version. It allows people to make the most of their (imbalanced) knowledge across two languages. 

\subsection{Measurements}

We employed three groups of measures in response to our hypotheses and RQs: participants’ subjective evaluation of the non-native speaking information seeker's agency (H1a, H1b), various objective measures of the dyad’s information exchange performance (RQ1a, RQ1b), and the perceived workload of the MT-assisted information-seeking task (RQ2).

\subsubsection{Non-native speakers' agency.} We used 7-point scales to collect each non-native speaker’s self-ratings and their native speaking partner’s ratings on their personal agency demonstrated in the task context. Survey items under this measurement were adopted from the Sense of Agency Scale \cite{tapal2017sense} (e.g., “[The item version rated by non-native speakers] I was in control of what my messages would convey to my partner,” or “[The item version rated by native speakers] My partner was in control of what their messages would convey to me;” Cronbach’s $\alpha$ = .80). A higher average rating indicated a higher level of agency perceived by the participant. 

\begin{small}
\begin{table*}[h]
\caption{Message categories and examples based on the information exchange functions they serve}
\label{tab:conv_code}
    \centering
    \begin{tabular}{>{\raggedright\arraybackslash}p{4.5cm}>{\raggedright\arraybackslash}p{8.5cm}}
    \toprule
    \textbf{Category and Description} & \textbf{Example Messages Within One Information Exchange }\\
\hline
\textbf{Content}: Initiation of a new information exchange
     & \textit{"House A and B are four rooms, but one is a single floor, one is a third floor. Which one would be easier.} \newline [Non-native speaker] \newline \\
      \textbf{Content}: Elaboration on an already-initiated exchange & 
        \textit{"That depends on how you like the stairs. Does anyone in your family have mobility issues?"}  \newline [Native speaker] \newline \newline 
        \textit{"The kids are 6 years old and 8 years old. I'm afraid they'll get hurt."} \newline [Non-native speaker]\newline
        \newline "So it's better on the single floor?" \newline [Non-native speaker] \newline
        \newline \textit{"That's a valid concern. When my children were that age we had a single story home but we still had to maneuver stairs to get into and out of the home. I'm not sure about either one of these homes as far as that is concerned"} \newline [Native speaker]\newline \\

 \textbf{Coordination}&  \textit{"Okay, I get it." }\newline [Non-native speaker]\\
 \bottomrule
\end{tabular}
\end{table*}
\end{small}

\subsubsection{Dyadic information exchange process.} The process of information exchange is commonly examined by its breadth (i.e., the number of unique information pieces exchanged in the conversation) and depth (i.e., the number of elaborations made on each unique information piece being exchanged). To apply these two measures in our task context, we manually coded all conversation logs recorded on the system server. Specifically, we adopted Clark’s classic framework, which categorized all messages in a conversation into two types \cite{clark1996using}: content-related messages that attempt to carry out the official task (e.g., the exchange of the immigrant family’s housing needs or a housing property’s characteristics) and coordination-related messages that do not contribute new content but are used to facilitate the conversation (e.g., “oh, okay” as a backchannel response). After that, we focused on each content-related message, categorizing it under one of the two groups: the initiation of a new information exchange, or the elaboration on an already-initiated exchange.  Table 2 detailed all the message categories used in this analysis, along with examples from our participants’ conversation logs. Two research assistants, blind to the rationale behind our study design, began by reviewing the entire corpus and independently coding a subset of all message turns. The inter-coder reliability achieved at the end of this step was already high, with Cohen's Kappa $\kappa$ = .82. Following this, the two assistants discussed their discrepancies, clarified their coding rules, and proceeded to code the remaining message turns. They repeated the calibration steps until they arrived at unified coding results for the entire corpus. Based on the coded corpus, we calculated the total number of initiation messages in each task conversation (i.e., the breadth), with each initiation message representing one unique information piece being exchanged. We also calculated the average number of elaboration messages per unique information piece (i.e., the depth), as each elaboration message further deepened the discussion surrounding its initial status. 

\subsubsection{Dyadic information exchange outcome.} Prior research on information exchange often considers the alignment between information seekers and information providers as a key outcome at the end of the exchange. We adopted this criterion to evaluate the dyadic outcome in our task context. Specifically, we included one question in the post-task survey, asking all participants to rank a shared list of five items according to their importance to the immigrant family’s housing selection. All five items on the list were derived from the task materials that participants had reviewed and centered their conversations around, including the anticipated financial expenses, the architectural condition of the house, the nearby services and amenities, the networking opportunities within the local community, and the educational resources available for children in the family. We computed the Rank-Biased Overlap (RBO) score to assess the alignment between the rankings provided by participants in the same dyad \cite{webber2010similarity}. This similarity score ranges between 0 and 1, with a higher value indicating greater alignment and, hence, better dyadic information exchange outcome. 

\subsubsection{Workload} We used 7-point scales to collect each participant’s self-ratings on the workload they experienced in the task context. Survey items under this measurement were adopted from the NASA TLX scale \cite{hart2006nasa} (e.g., “How much mental burden (e.g., thinking, searching, deciding) did you experience during the entire conversation?”; Cronbach’s $\alpha$ = .66). A higher average rating indicated a higher level of workload perceived by the participant.  

\section{RESULTS}
We conducted a series of ANOVAs to examine the effect of the human-MT interface design on several factors at the center of our interest: the preservation of non-native speakers’ agency in MT-assisted conversations (H1a, H1b), dyadic performance as revealed in participants’ information exchange process (RQ1a) and outcome (RQ1b), and the perceived task workload (RQ2). We applied the Bonferroni correction to all pairwise comparisons. The remainder of this section details the model we used in each analysis and the relevant results. 

\subsection{Non-Native Speakers' Agency}
\begin{figure}[h]
    \centering
    \includegraphics[width=1\linewidth]{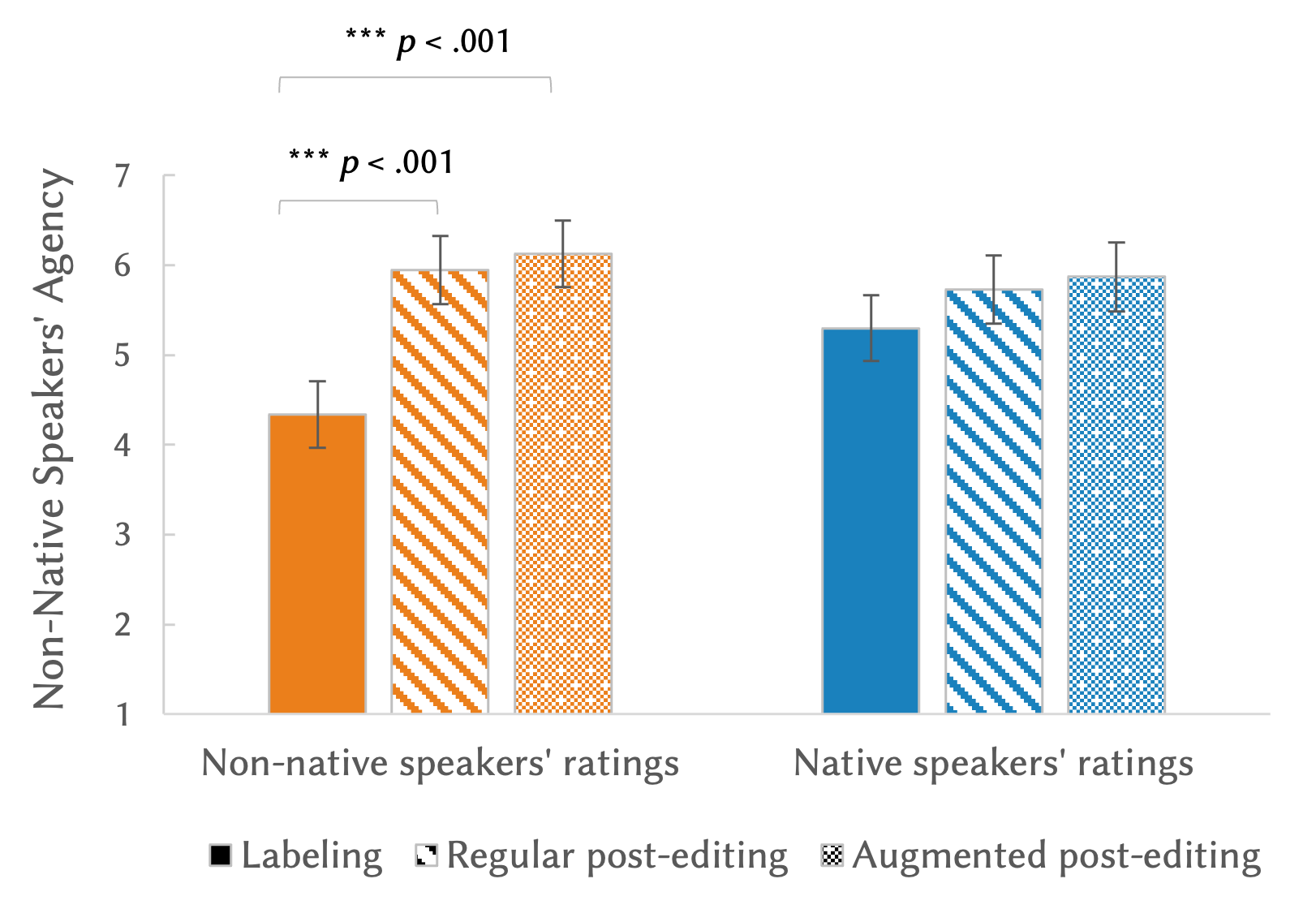}
    \caption{Non-native speakers' agency by the human-MT interface condition and the rater's language background}
    \label{fig:NNS_agency}
    \Description{This bar chart displays non-native speakers' agency by the human-MT interface condition and the rater's language background. When the raters were non-native speakers themselves, they gave significant higher rating in the augmented post-editing condition than the labeling condition (p < .001) and higher rating in the regular post-editing condition than the labeling condition (p < .001). When the raters were native speakers, there were no significant difference in their rating across the three conditions.}
\end{figure}

Our H1a and H1b predicted that the three human-MT interfaces would all help non-native speakers sustain their personal agency in MT-assisted conversation, but the level would vary across conditions. We performed a 3 $\times$ 2 Mixed Model ANOVA to test the hypotheses. The results of this analysis supported H1a and rejected H1b (\autoref{fig:NNS_agency}).

Specifically, we set the dependent variable of this analysis to be each participant’s rating of the non-native speaker’s agency exercised in the task conversation. One independent variable was the human-MT interface condition: labeling interface, regular post-editing interface, or augmented post-editing interface. A second independent variable was the rater’s language background: non-native speaker or native speaker. Participants were nested within the dyads. The Satterthwaite’s approximation was applied to estimate the degrees of freedom, which often generated non-integer values. Control variables in this analysis included each person’s demographic information (e.g., age, gender), existing knowledge about housing selection in the United States, prior experience with cross-lingual communication with people of the other residence status, and prior experience in using language technology. 

The results indicated a significant main effect of the human-MT interface condition: F [2, 41.43] = 8.17, \textit{p} < .001. The main effect of language background was not significant: F [1, 72.68] = 0.09, \textit{p} = .76. The interaction effect between these two factors was marginally significant: F [2, 40.19] = 3.19, \textit{p} = .05. 

We used pairwise comparisons to examine ratings provided by the non-native speakers themselves and the native speakers working with them, respectively. In support of H1a, non-native speakers rated themselves as having a higher level of agency in the regular post-editing interface condition (M = 5.95, S.E. = .38) than in the labeling interface condition: F [1, 77.67] = 14.99, \textit{p} < .001. They also reported a higher level of agency in the augmented post-editing interface condition (M = 6.13, S.E. = .37) than in the labeling interface condition (M = 4.34, S.E. = .37): F [1, 77.82] = 18.43, \textit{p} < .001. That said, there was no significant difference between the two post-editing conditions, rejecting H1b. Meanwhile, the native speakers provided similar ratings to the non-native speakers’ agency across the augmented post-editing interface condition (M = 5.87, S.E. = .38), regular post-editing interface condition (M = 5.73, S.E. = .38), and labeling interface condition (M = 5.30, S.E. = .37). 

In summary, the above results suggested that the non-native speakers' self-perceived agency tended to vary according to the influence they could make on MT outputs: the agency was higher in the two post-editing interface conditions, compared to the labeling interface condition. This difference, however, was not observed in ratings provided by their native speaking partners.

\subsection{Dyadic Information Exchange Process}
\begin{figure*}[h]
    \centering
    \includegraphics[width=1\linewidth]{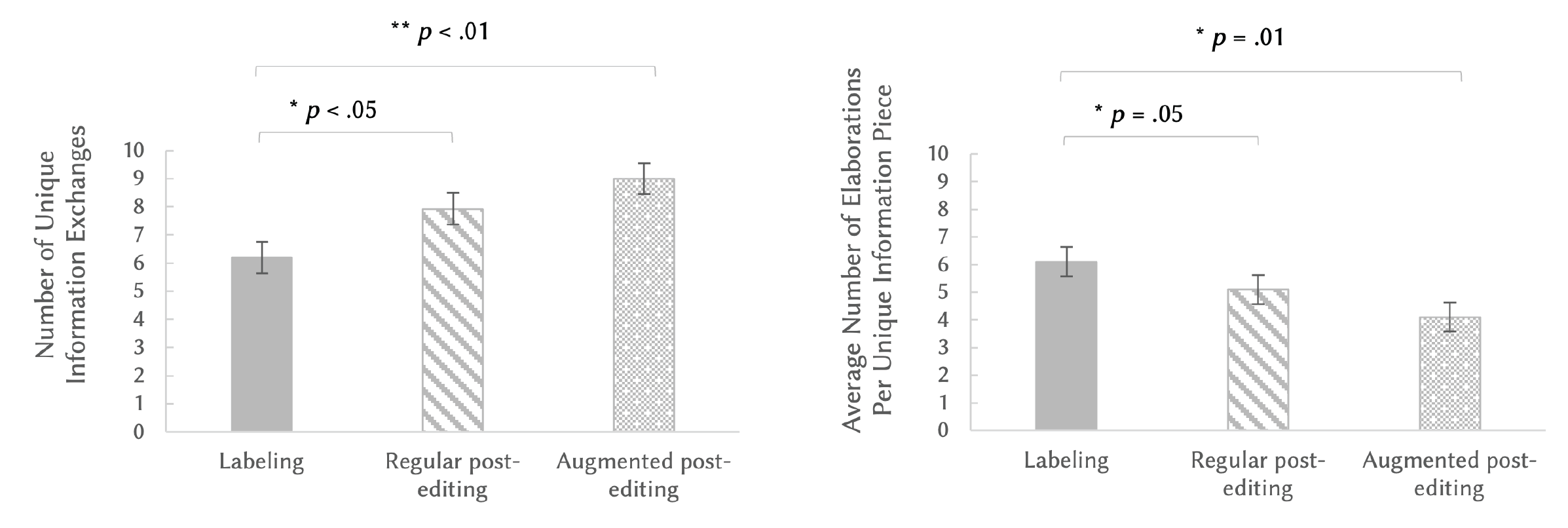}
    \caption{Dyadic information exchange process as in the breadth (left) and depth (right) by the human-MT interface condition}
    \label{fig:process}
    \Description{Figure 4 includes two bar charts that display the dyadic information exchange process as in the breadth (left) and depth (right) by the human-MT interface condition. The left chart indicates that breadth was significantly higher in the augmented post-editing condition than the labeling condition (p < .01) and higher in the regular post-editing condition than the labeling condition (p < .05). The right chart indicates that depth is lower in the augmented post-editing condition than the labeling condition (p = .01) and lower in the regular post-editing condition than the labeling condition (p = .05).}
\end{figure*}

We wondered how the adoption of the three human-MT interfaces would affect participants' dyadic information exchange process (RQ1a). In the context of this research, we examined the process by analyzing the breadth and depth of our participants’ information exchange. Overall, our analyses showed that both the breadth and the depth of information exchange varied according to the human-MT interface condition, but these variations occurred in two opposite directions (\autoref{fig:process}).

\subsubsection{Breadth of information exchange} We used a one-way ANOVA to compare the total initiations of unique information pieces exchanged by dyads across three human-MT interface conditions. The dependent variable was the total number of unique information exchanges initiated per dyad. The independent variable was the human-MT interface condition: labeling interface, regular post-editing interface, or augmented post-editing interface. Control variables were the same as those used for agency. The results indicated a significant main effect of the human-MT interface condition: F [2, 42] = 6.32, \textit{p} < .005. Specifically, dyads of participants initiated a lower total number of unique information exchanges under the labeling interface condition (M = 6.20, S.E. = .56) than under the regular post-editing interface condition (M = 7.93, S.E. = .56; F [1, 42] = 4.75, \textit{p} < .05 ) or the augmented post-editing interface condition (M = 9.00, S.E. = .56; F [1, 42] = 12.39, \textit{p} < .01). There was no significant difference between the two post-editing interface conditions. 

\begin{figure}[h]
    \centering
    \includegraphics[width=1\linewidth]{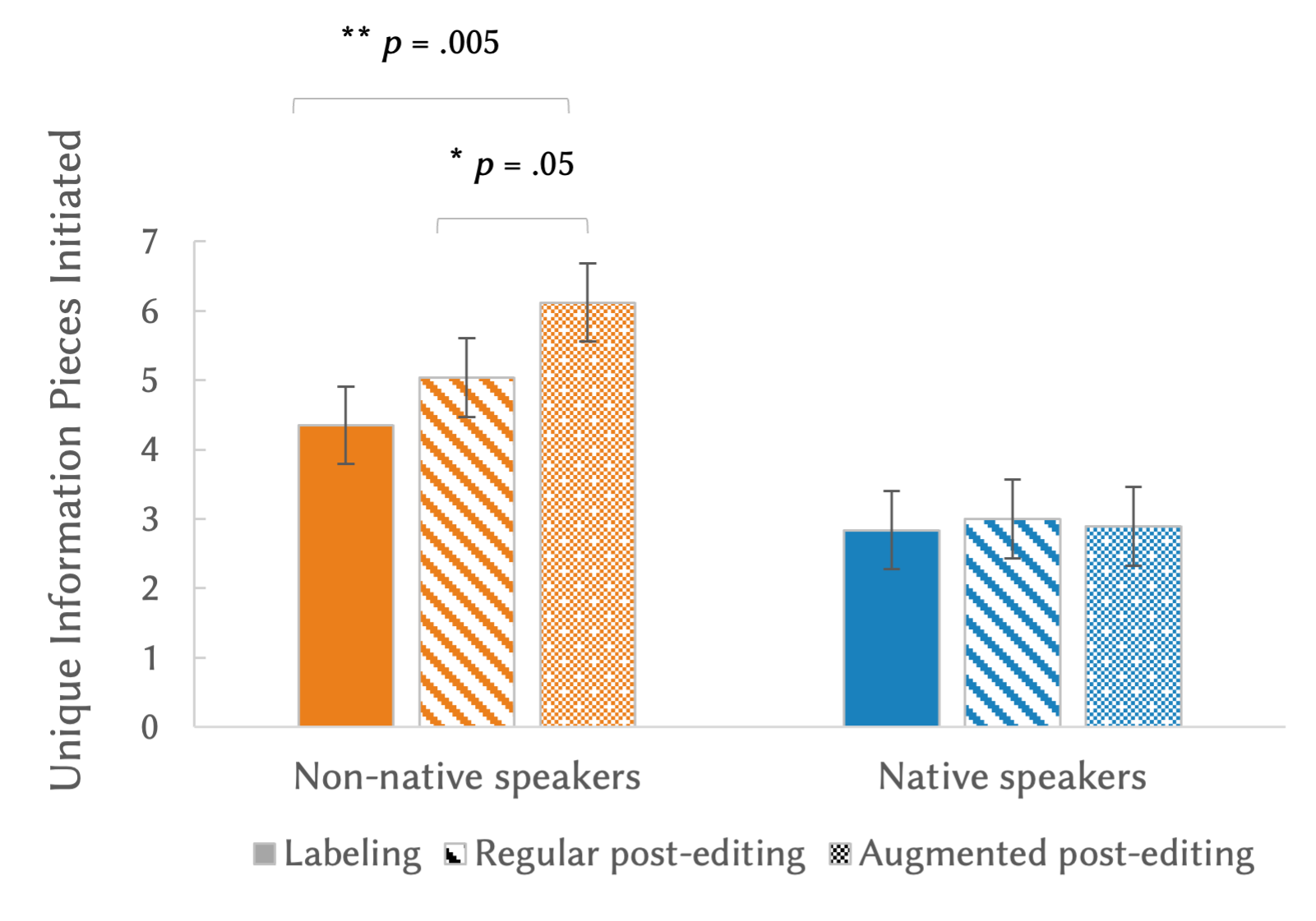}
    \caption{Unique information pieces initiated by the human-MT interface condition and individual language background}
    \Description{Figure 5 shows the number of unique information pieces initiated by the human-MT interface condition and individual language background. For non-native speakers, they initiated more unique information pieces in the augmented post-editing condition than the regular post-editing condition (p = .05) and more unique information pieces in the augmented post-editing condition and the labeling condition (p = .005). For native speakers, there were no significant difference in the number of unique information pieces initiated across the human-MT interface conditions.}
    \label{fig:Breadth_individual}
\end{figure}

To further unpack this result, we also compared the total number of unique information pieces that the non-native speakers and the native speakers had initiated in their conversations, respectively. The analysis was performed with a 3 × 2 Mixed Model ANOVA (\autoref{fig:Breadth_individual}), where the dependent variable was the total number of unique information exchanges initiated per person. The rest of the model setup was the same as that used for agency. The results indicated a significant main effect of the human-MT interface condition: F [2, 40.46] = 6.92, \textit{p} < .005. The main effect of language background was significant as well: F [1, 78.87] = 10.81, \textit{p} < .005. There was no significant interaction effect between these two factors: F [2, 41.46] = 0.87, \textit{p} = .43. 

Pairwise comparisons revealed that non-native speakers contributed more to the initiation of unique information exchanges under the augmented post-editing interface condition (M = 6.12, S.E. = .56) than in the regular post-editing interface condition (M = 5.04, S.E. = .57; F [1, 74.07] = 3.42, \textit{p} = .05.) or the labeling interface condition (M = 4.35, S.E. = .56; F [1, 74.94] = 8.34, \textit{p} = .005). In contrast, native speakers’ contributions to the initiation of unique information exchanges appeared similar across the augmented post-editing interface condition (M = 2.89, S.E. = .57), the regular post-editing interface condition (M = 3.00, S.E. = .57), and the labeling condition (M = 2.84, S.E. = .56). 

\subsubsection{Depth of information exchange. }We used another one-away ANOVA to compare the average number of elaborations in dyadic conversations across three human-MT interface conditions. The dependent variable was the average number of elaborations generated per dyad. The rest of the model setup was the same as that used for dyadic breadth of information exchange. The results indicated a significant main effect of the human-MT interface condition: F [2, 42] = 30.09, \textit{p} < .05. Specifically, dyads of participants made a larger average number of elaborations under the labeling interface condition (M = 4.10, S.E. = .53) than under the regular post-editing interface condition (Mean = 5.10, S.E. = .53; F [1, 42] = 1.98, \textit{p} = .05) or the augmented post-editing interface condition (M = 6.11, S.E. = .53; F [1, 42] = 7.23, p = .01).  There was no significant difference between the two post editing conditions.

\begin{figure}[h]
    \centering
    \includegraphics[width=1\linewidth]{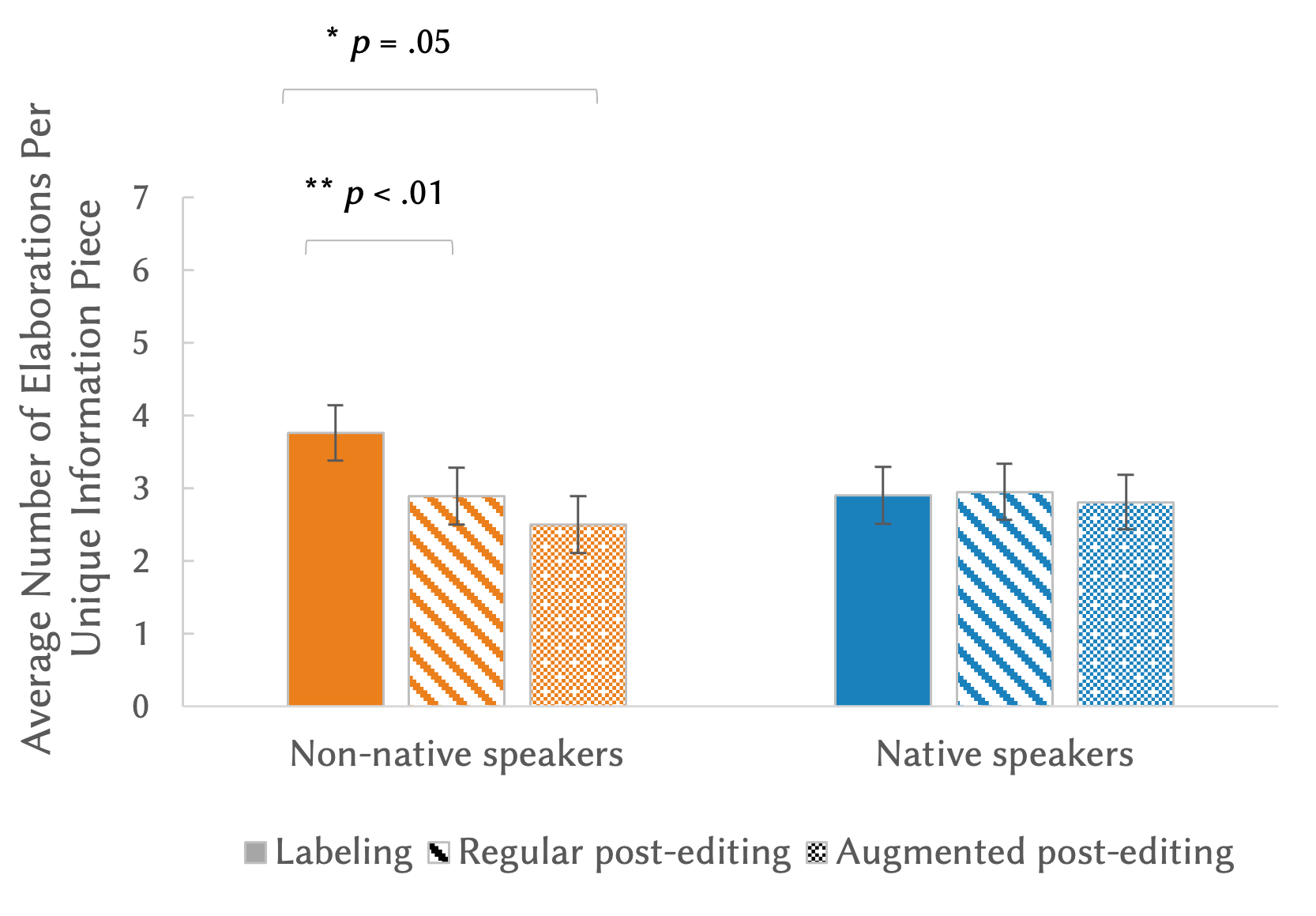}
    \caption{Elaborations added by the human-MT interface condition and individual language background}
    \label{fig:Depth_individual}
    \Description{Figure 6 shows the average number of elaborations by the human-MT interface condition and individual language background. For non-native speakers, they added significantly more elaborations in the labeling condition than the regular post-editing condition (p < .01) and the augmented post-editing condition (p = .05). For native speakers, there were no significant difference in the number of elaborations across the human-MT interface conditions.}
\end{figure}

To gain further insights, we compared the average number of elaborations that non-native speakers and native speakers had added to their conversations, respectively. The analysis was performed with a 3 × 2 Mixed Model ANOVA (\autoref{fig:Depth_individual}), where the dependent variable was the average number of elaborations generated per person. The rest of the model setup was the same as that used for agency. The results indicated no significant main effect of the human-MT interface condition (F [2, 41.77] = 1.34, \textit{p} = .27), nor of the language background (F [1, 53.65] = 0.13, \textit{p} = .72). However, there was a significant interaction effect between these two factors: F [2, 38.35] = 4.29, \textit{p} < .05. 

Pairwise comparisons revealed that non-native speakers contributed more to the generation of elaborations under the labeling interface condition (M = 3.76, S.E. = .38) than in the regular post-editing interface condition (M = 2.89, S.E. = .39; F [1, 61.61] = 7.34, \textit{p} < .01) or the augmented post-editing interface condition (M = 2.50, S.E. = .39; F [1, 61.28] = 3.56, \textit{p} = .05). In contrast, native speakers’ contributions to the elaborations turned out to be equal across the augmented post-editing interface condition (M = 2.81, S.E. = .39), the regular post-editing condition (M = 2.95, S.E. = .39), and the labeling condition (M = 2.90, S.E. = .38). 

In short, the above results suggested that the dyadic information exchange process varied according to the influence non-native speakers had on MT outputs: conversations under the labeling interface condition featured the highest level of depth but the lowest level of breath. No significant differences were found between the two post-editing interface conditions. More interestingly, non-native speakers appeared to act as the primary drivers of those observed differences, while native speakers performed similarly across all three interface conditions. 

\subsection{Dyadic Information Exchange Outcome}

\begin{figure}[h]
    \centering
    \includegraphics[width=1\linewidth]{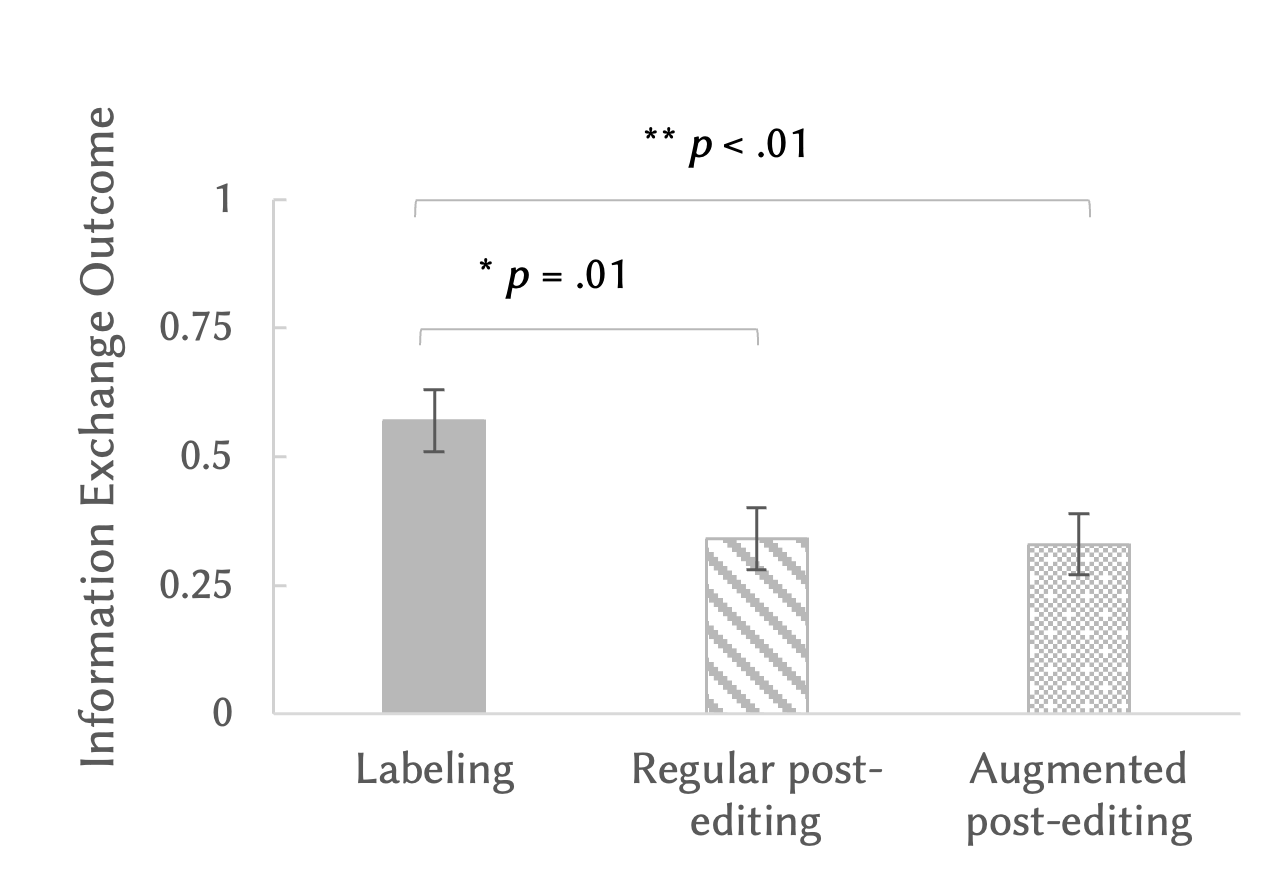}
    \caption{Dyadic information exchange outcome by the human-MT interface condition}
    \label{fig:outcome}
    \Description{Figure 7 shows the dyadic information exchange outcome by the human-MT interface condition. Dyads in the labeling condition had significantly better outcome than the regular post-editing condition (p = .01) and the augmented post-editing condition (p < .01).}
\end{figure}

Our RQ1b asked how the adoption of the three human-MT interfaces would affect participants' dyadic information exchange outcome. To answer this question, we conducted a one-way ANOVA to compare the dyadic ranking alignment scores across three human-MT interface conditions (\autoref{fig:outcome}). The dependent variable was the level of alignment between the two participants within a dyad. The rest of the model setup was the same as that of the dyadic breadth of information exchange. The results indicated a significant main effect of the human-MT interface condition. Specifically, dyads achieved a greater level of alignment when their information exchange took place under the labeling interface condition (Mean = .57 S.E. = .06) compared to the regular post-editing interface condition (Mean = .34 S.E. = .06; F [1, 42] = 6.74, \textit{p} = .01) or the augmented post-editing interface condition (Mean = .33 S.E. = .06; F [1, 42] = 7.55, \textit{p} < .01). There was no significant difference between the two post-editing interface conditions. 

In short, the dyadic task outcome, in terms of alignment achieved between two parties of the information exchange, was best under the labeling interface condition.  

\subsection{Workload }

Our RQ2 asked how the adoption of the three human-MT interfaces would affect participants’ perceived workload. We performed a 3 × 2 Mixed Model ANOVA to answer this question. The model setup was the same as that used for agency, only except that the dependent variable became each participant’s self-rating of their workload during the task. The results demonstrated no difference in workload across the three interface conditions (F [2, 41.32] = 0.44, \textit{p} = .65), nor between participants with different language backgrounds (F [1, 77.41] = 0.97, \textit{p} = .33). The interaction effect was not significant either (F [2, 41.24] = 0.20, \textit{p} = .82). 

That is, the perceived workload was consistent across all three interface conditions and for all participants.

\section{DISCUSSION}

To recap, the conventional design of human-MT interface today typically follows a straightforward two-box layout: non-native speakers of a target language (e.g., English) type in their source message in one text box; they receive and may preview the translated version of that message in another box before sending. Such an interface design generally does not allow users to exercise their agency by making further influence on the MT outputs. Our current work investigated ways to sustain non-native speakers’ agency in MT-mediated conversations. We explored three potential designs for the human-MT interface. Each interface allowed non-native speakers to leverage their resources to a distinct extent, thereby influencing the translated messages sent to their task partners.

In the remainder of the Discussion section, we reflect on how our findings relate to existing literature in HCI and beyond. These reflections focus on three key aspects of our results: the agency-resource connection (Section 5.1), the associated costs of sustaining non-native speakers’ agency (Section 5.2), and the comparable levels of agency observed under two post-editing conditions (Section 5.3). We conclude this section by presenting design insights derived from these reflections (Section 5.4).

\subsection{Operationalizing Agency Preservation Through the Agency-Resource Connection}

We speculated three conditions of human-MT interface for satisfying non-native speakers’ agentic needs: the labeling condition, the regular post-editing condition, and the augmented post-editing condition. In each case, the average self-ratings given by non-native speakers confirmed that resource involvement would effectively anchor personal agency in MT-mediated conversations. Participants reported preserving medium to high levels of their personal agency when working with all three tested interfaces. Across conditions, the comparison of non-native speakers’ self-ratings revealed how personal agency would vary based on the scope of resources they leveraged to influence MT outputs. Participants reported significantly lower agency when they utilized only a target portion of their language resources to label the quality of MT outputs, compared to when they used a broader set of their language resources for direct editing.

The above findings highlight the unique value of adopting Bandura’s theoretical lens to better sustain non-native speakers’ agency. In particular, Bandura’s theory underscores the importance of understanding personal agency within the context of resource involvement. Beyond understanding, it also calls for adjusting the agency an individual can excise by manipulating the resources they can access \cite{bandura1999, bandura2006toward, banduraSelfEfficacyMechanismHuman1982, banduraPsychosocialImpactMechanisms2003}. Such a perspective not only provides HCI researchers with conceptual insights into the behavioral foundation of personal agency but also suggests actionable strategies for moderating agency through system interface design.

Inspired by the agency-resource connection, we suspect that the regular post-editing interface, commonly used to preserve the agency of fluent bilinguals (e.g., professional translators; \cite{cadwell2016human, guerberof2013professional, koponen2016machine, obrienHowDealErrors2022}), may not be the sole reasonable option to sustain non-native speakers’ agency. We specified and compared three human-MT interfaces that enabled non-native speakers to utilize different scopes of their language resources in a progressive manner. The labeling interface addresses non-native speakers’ agentic needs by leveraging their passive English vocabulary – a limited but reliable subset of the person’s language resources – for recognizing or evaluating the quality of MT outputs. The augmented post-editing interface, conversely, served the same needs by activating more of the non-native speakers’ active English vocabulary to facilitate their direct editing of MT outputs. Together, these interfaces expand our view on how to preserve a non-native speaker’s agency, taking into account the sharper divide between passive and active vocabulary in the composition of these individuals’ language resources in English.

Notably, while we adopted Bandura’s agency-resource connection to guide the current investigation of agency, our intention has never been to diminish the importance of other relevant theories. In fact, the concept of agency has been widely discussed in numerous previous HCI projects. According to Bennett et al.’s recent review \cite{bennettHowDoesHCI2023}, a substantial body of this prior work draws on self-determination theory to support their interpretations of agency. Other relevant frameworks include Suzy Killmister’s four dimensions of self-governance \cite{killmister2017taking, killmister2013autonomy}, Frankfurt’s hierarchical notion of freedom \cite{frankfurt2018freedom, frankfurt2003lternate}, and many others (e.g., \cite{ryan2000self, emirbayer1998agency, dworkin1988theory, sullivan1989immanuel}). Most of these theories emphasize the importance of an individual’s self-identity, personal values, intrinsic motivations, and desires in shaping their view of agency. Compared to them, Bandura’s theory focuses more on an individual’s exercise of agency \cite{bandura1999, bandura2006toward, banduraSelfEfficacyMechanismHuman1982, banduraPsychosocialImpactMechanisms2003}, often through the use of their resources as well as the task environment that contextualizes the access to and utilization of such resources. This comparison makes Bandura’s framework particularly suitable for the purposes of our research, which centers on ideating and operationalizing human-MT interface design for non-native speakers’ agency preservation.

\subsection{Attending to Associated Costs of Non-Native Speakers' Agency}

The results of our study demonstrated the potential of all three tested human-MT interfaces in supporting non-native speakers' agency. This naturally raises the question: Does one of these interfaces perform significantly better or worse than the others? Should we, as researchers and designers, recommend that future MT-mediated systems adopt or avoid any of the particular interfaces tested here? Reflecting on these questions, it becomes clear that we must first define what qualifies as a "better" or "worse" human-MT interface.

If we consider the preservation of non-native speakers' agency as the sole criterion, we might conclude that the labeling interface is less effective than the two post-editing interfaces. While such an interface design supported participants' agency, it did so only to a moderate degree. However, this conclusion overlooks other critical aspects of dyadic communication, especially when such communication serves high-stakes purposes, such as facilitating essential information seeking for immigrants. 

To gain a more comprehensive understanding, we analyzed additional measures to explore how dyadic information exchange performance varied across the different human-MT interfaces provided to non-native speakers. Our findings revealed that the variation in agency had interesting associated effects. In the labeling condition, dyads exchanged fewer unique pieces of information compared to other conditions. However, they engaged in deeper discussions about each piece of information. Our data indicated that non-native speakers were the primary drivers of these in-depth discussions. It also revealed more successful alignment between the two speakers by the end of their information exchange.

Thus, when we adopt a holistic perspective that considers multiple qualifiers for successful communication, important strengths of the labeling interface become evident. The boosts in non-native speakers’ agency observed under the two post-editing conditions came at significant costs. One possible explanation for these costs is that utilizing active English vocabulary to edit MT outputs depleted the attention participants might otherwise have allocated to probing further into their discussions with task partners. Similar tensions have also been reported in recent literature on AI-mediated communication (e.g., \cite{buschek2021impact, fu2024text, kim2024diarymate}), and they may be even more pronounced for non-native speakers, whose language resources are characterized by the “recognition and production gap \cite{gass1983development}.”

Given these reflections, we believe that our study contributes an empirical understanding of the costs associated with supporting non-native speakers' agency. It complements existing HCI literature by addressing a critical shortfall: while there is broad consensus on the value of supporting agency, prior research often fails to clarify how agency relates to other aspects of users' experiences within specific task environments \cite{bennettHowDoesHCI2023}.

Returning to the question of what constitutes a "better" or "worse" human-MT interface, we find no straightforward answer due to the associated costs of agency, as detailed above. When the priority is to understand the detailed reasoning behind an immigrant’s information needs, the labeling interface is a good choice: it maintains a degree of personal agency in language use for non-native speakers and facilitates in-depth information exchange between the immigrant and the local person. However, when the priority shifts to understanding the landscape of an immigrant’s information needs without delving into details during a single conversation, the two post-editing interfaces may be more effective.

We believe it is important to raise awareness, within the research community as well as among MT users, of the potential trade-offs between agency and other aspects of communication. Equipped with this understanding, people can make informed decisions about human-MT interface designs that align with their priorities in specific contexts.

\subsection{Making Sense of the Comparable Levels of Agency Under Two Post-Editing Conditions}

The comparison of non-native speakers’ agency between two post-editing conditions did not support our H2b. While the LLM-generated paraphrases were designed to activate more of the active English vocabulary of non-native speakers and scaffold their post-editing, this did not result in higher agency in the augmented condition, compared to the regular condition. 

We uncovered additional evidence from the data to make sense of the above result. We examined all conversation logs from the augmented post-editing condition to verify whether non-native speakers in our study have utilized these language resources. Our data showed that, on average, participants consulted the paraphrases in 30.3\% (S.E. = 9.1\%) of the messages they sent. They adopted words and expressions from these paraphrases in 19.6\% (S.E. = 6.4\%) of their messages. Among all messages edited in the augmented post-editing condition, an average of 62.7\% (S.E. = 10.1\%) of the edited tokens were adopted from the LLM-generated paraphrases. This data confirms that the language resources hinted via the augmented interface did facilitate our participants’ use of their active English vocabulary during MT output editing.

Thus, rather than suspecting a lack of engagement with the LLM-generated resources or a failure in establishing the agency-resource connection, we believe a “ceiling effect” provides a more reasonable explanation for our findings. That is, non-native speakers’ agency in the regular post-editing condition was already high (i.e., an average score of 5.95 out of 7), leaving limited room for another significant increase in the augmented condition.

Moreover, some measures analyzed in our study highlighted how the augmented post-editing interface benefited our participants, although this benefit was not directly related to agency. Specifically, we observed that non-native speakers in the augmented post-editing condition initiated more unique information exchanges with their task partners compared to how they behaved in other conditions. This finding, along with those discussed in Section 5.2, underscores the necessity of remaining mindful of associated costs and benefits when valuing personal agency.

\subsection{Deriving Insights for Sustaining Non-Native Speakers' Agency Through System Design}

Building on the critical aspects of the results discussed above, we outline three key directions for sustaining the agency of non-native speakers. These directions extend beyond the specific task context of this study. The first two focus on future human-MT interfaces to support conversational interactions, while the third envisions how such interfaces could facilitate innovative data collection methods for training translation algorithms.

\subsubsection{Labeling as a low-cost and valuable feature for conversational interfaces}

Our analysis underscores the potential of the labeling interface to support high-quality dyadic communication while preserving the personal agency of non-native speakers at a moderate level. Notably, similar labeling functions have already been integrated into several commercial MT systems. For instance, Google Translate and DeepL both encourage users to evaluate the translations they receive by clicking a thumbs-up or thumbs-down label. While these interfaces are convenient to use, they are not designed for real-time conversational contexts. The labels are primarily utilized by developers to improve translation services in the long term, rather than enabling users to influence MT outputs or communicate their inputs to conversation partners during a real-time interaction.

Some recent studies on MT usage have implemented labeling interfaces, similar to the one discussed in our work, for participants to flag low-quality translations encountered in MT-mediated instant message exchanges \cite{robertson2022understanding}. However, the focus of that research was not on examining personal agency during communication but rather on capturing human perceptions of translation quality.

Building on these prior efforts, our work emphasizes the broader value of incorporating the labeling feature into text-based communication tools equipped with MT modules. This feature is especially helpful for non-native speakers with low proficiency in the target language. It provides them with a feasible mechanism to effectively engage with a subset of their language resources (e.g., passive vocabulary in the target language), thereby exercising their influence on the messages sent to conversational partners.

\subsubsection{Promoting awareness of the costs associated with enhanced agency}

Besides incorporating the labeling function, our study results suggest other, more straightforward ways to enhance non-native speakers’ agency in MT-mediated conversations. For instance, the interface could let the translated text in the message box remain editable until the communicant clicks the "send" button. This feature would be easy to implement and holds promising potential to encourage non-native speakers to post-edit their messages. However, many instant messaging tools today, such as the messenger module on all versions of iPhones, have yet to consider this feature.

A substantial part of our results revealed costs that may be associated with enhanced agency in post-editing conditions. Further analysis pointed to reasons for these costs: while non-native speakers may attempt to utilize their active English vocabulary to edit MT outputs, their limited English proficiency makes this process effort-intensive. Consequently, post-editing appeared to distract these individuals from engaging in in-depth information exchanges with their conversational partners. To address this constraint, we envision that future system designs could encourage non-native speakers to post-edit MT outputs selectively, focusing on moments where high-stakes miscommunication is likely to occur. This guidance on prioritizing post-editing efforts may help users balance the benefits and costs of using a post-editing interface.

Recent advancement in LLMs provides the technical feasibility to generate the described guidance. In particular, several research projects have demonstrated the success of using LLMs to identify selected discourse acts \cite{stolcke2000dialogue, ritter2010unsupervised, niederhoffer2002linguistic, soler2023measuring}, track various aspects of the conversation dynamics over time \cite{mao2020dialoguetrm, hua2024did}, and detect interaction patterns that reveal miscommunications \cite{higashinaka2016dialogue, kementchedjhieva2021dynamic, finch2023leveraging, raheja-etal-2023-coedit, ziegenbein-etal-2024-llm}. These techniques could be applied to highlight moments where post-editing is most crucial for communicants. 

\subsubsection{Building customized MT systems based on first-person account data}

The labeling and post-editing interfaces explored in this study open up new technical opportunities to train customized MT algorithms that reflect an individual’s word choices and language styles, thereby sustaining their agency in a proactive manner. They hold the promise of offering complementary information to that used in today’s MT system-building practices. The latter often relies on training data generated by third-party annotators (e.g., crowd workers who review the text corpus resulting from others’ conversations), without incorporating first-person perspectives contextualized to the actual communicant \cite{agrawal-etal-2024-automatic-metrics}. 

More specifically, labels generated by non-native speakers during their real-time communication could serve as reinforcement learning feedback to personalize MT models \cite{saluja-etal-2012-machine, michel-neubig-2018-extreme, kreutzer-etal-2020-correct, naradowsky-etal-2020-machine, agrawal-etal-2024-modeling}. Developers could also train a classifier to predict the same person’s preferences in follow-up conversations over time  \cite{ziegler2019fine}. Post-edits collected through the human-MT interfaces tested in our study could serve as in-context learning examples, aiding the fine-tuning of translation models \cite{bhattacharyya-etal-2022-findings, bhattacharyya-etal-2023-findings, occhipinti-etal-2024-fine, moslem-etal-2023-adaptive, agrawal-etal-2023-context} and correcting tacit errors deemed important by a given communicant \cite{treviso-etal-2024-xtower, ki-carpuat-2024-guiding}. More collaborative research between HCI and NLP experts may emerge as more data from the first person account becomes available.

\section{LIMITATIONS AND FUTURE DIRECTIONS}

As one of the first HCI studies to examine non-native speakers’ agency in MT-mediated conversations, our work was conditioned by the specific methodological choices and task settings described in this paper. Below, we reflect on the composition of participants, the three human-MT interfaces, and the communication modalities examined in this work. We detail our thoughts on the generalizability of our findings, as well as opportunities for future research.

\subsection{The Composition of Participants}

We investigated ways to sustain non-native speakers’ agency in the context of essential information-seeking interactions between Mandarin-speaking immigrants (as seekers) and English-speaking local residents in the United States (as providers). This choice was motivated partly by the real-world language challenges we observed through communal work with this immigrant group and partly by the linguistic differences between Mandarin and English, which are of empirical significance for MT research. For example, the syntactic and pragmatic distance between Mandarin and English is much greater than that between most Romance languages and English. In the former case, it becomes even more critical to encourage non-native speakers’ exercise of their personal agency in monitoring, evaluating, and influencing MT outputs.

If we were to switch to other immigrant groups speaking different native languages, as long as they have imperfect proficiency in English, the “recognition and production gap” in their use of English language resources would persist. Thus, our proposal to leverage the agency-resource connection as a conceptual foundation for preserving these individuals’ agency would remain applicable.

In addition to native language type, English fluency level constitutes another important aspect of a non-native speaker’s linguistic profile. Participants in our sample exhibited a medium level of English fluency, with small individual differences in this measure. The broader population of non-native speakers includes individuals with both lower and higher English fluency than our participants. In such cases, we believe that the overall design approach of enabling resource involvement at varying scopes would still help sustain people’s agency in MT-mediated conversations. The more constraints a non-native speaker faces in utilizing their active vocabulary, the more likely they are to favor and benefit from the labeling interface. 

Alternatively, these individuals could be provided with an augmented post-editing interface, where additional hints are offered to help activate more of their own language resources in English. A potential issue with this second setup is that it may be cognitively taxing for individuals to process the hints provided via the interface. It could raise more serious concerns regarding the associated costs discussed in our study, such as the reduced quality of the dyadic information exchange. Given these considerations, we suspect that the labeling interface can support agency preservation across a broader range of non-native speakers, whereas the post-editing interfaces may not offer the same level of accessibility or benefit.

\subsection{The Three Human-MT Interfaces}

Inspired by Bandura’s agency-resource connection, we speculated on three human-MT interface designs. Our empirical comparisons across these interfaces showcased two pathways for calibrating a person’s resource involvement to exercise agency. One pathway engages a smaller-than-full scope of a person’s existing resource pool, as evidenced by our analysis and discussion of the labeling condition. The other provides additional hints to scaffold the activation of a person’s full resources, as demonstrated in the augmented post-editing condition. They highlight two generalizable approaches for designing human-MT interfaces that meet the agentic needs of non-native speakers.

For the first pathway, in particular, we do not yet have well-developed ideas for other interface designs that also enable the involvement of a non-native speaker’s passive vocabulary but do not follow the format tested in our study. We anticipate that readers of our paper may perceive the labeling condition as somewhat different from the two post-editing conditions: in the labeling condition, each message received by native speakers consisted of MT-generated textual content and labels provided by non-native speakers. In the two post-editing conditions, each message received by native speakers consisted of the MT-generated content and any edits made by non-native speakers; however, these two components appeared more visually integrated. 

We were less concerned about this difference for two reasons. First, all native speakers in our study were made aware that their conversational partners were adding influence, through labeling or editing, to those translated messages before sending them (see Section 3.2.2 for details). Second, the agency ratings collected from native speakers showed no significant difference across conditions, suggesting that native speakers perceived their conversational partners’ agency at an equal level, regardless of the human-MT interface used by those partners. For non-native speakers, even after assigning more visually prominent labels to MT outputs, they still rated their agency lower in the labeling condition than in other conditions. Such results align with what Bandura’s theory would predict, indicating the comparative benefits and limitations of using the labeling interface to support people’s agentic needs.

For the second pathway, we envision that numerous alternative interface designs could be explored beyond the augmented interface. For instance, the number of paraphrases and the format of LLM-generated language resources could both be adjusted. The interface could also provide clues to assist non-native speakers' critical assessment and adoption of AI-generated resources, such as translation quality feedback (e.g., \cite{mehandru-etal-2023-physician}) and structured prompts to aid people's critical thinking process (e.g., \cite{buccinca2021trust, he2024err}).


\subsection{The Communication Modality}

Our current work focuses on text-based conversations. This setup captures the translation functions used in available instant messaging applications, such as Slack, WeChat, and iMessage. The text-based modality enables individuals to interact with MT outputs and seek paraphrases for each written message. It also allows the researcher to log and track the entire conversation. That said, these benefits may not fully extend to speech-based translations, where additional challenges for non-native speakers’ language use often occur (e.g., \cite{hara2015effect}). Future work should explore how to sustain human agency in cases involving variations in participant composition, human-MT interfaces, and communication modalities beyond those examined in our study. The research materials and findings from our current study pave the way for this exciting area of exploration.

\section{CONCLUSION}

The current paper presents an experimental investigation involving 45 dyads, aimed at exploring the design of human-machine translation (MT) interfaces to sustain non-native speakers' agency in MT-assisted conversations. Each dyad was tasked with exchanging information in a given real-life scenario. In this scenario, one non-native speaker, acting as an immigrant information seeker, sought housing selection tips and suggestions for their family in the United States. One native English speaker, acting as a local information provider, engaged in the conversation with the non-native speaker to assist with their needs. Inspired by Bandura's notion of the agency-resource connection, we proposed three different human-MT interfaces and provided each dyad with one of them. Interestingly, we found that the labeling interface – the one designed to help non-native speakers maintain a moderate level of agency – facilitated dyadic information exchange with better performance. The two post-editing interfaces could sustain a higher level of non-native speakers' agency but at the cost of dyadic performance. These results highlight the complexity of human agency in MT-assisted conversations. In particular, they underscore the critical trade-off between personal agency and collective performance in non-native speakers' use of MT. Based on these findings, we outline potential directions for future human-MT interface design that can preserve non-native speakers' agency while also guiding people to remain mindful of the associated costs.

\begin{acks}
This research was supported by the National Science Foundation, under grants \#2147292 and \#2229885. We thank the reviewers for their constructive feedback on this work. We are also grateful to Ruipu Hu, Connie Siebold, Siyi Zhu, Elana Blinder, Yongle Zhang, Jian Zheng, Victoria Chang, and Yuhang Zhou for their insights during the early stages of our system development and research design. Last but not least, we thank all participants who engaged in our study and shared their task experiences with us.
\end{acks}



\appendix
\section*{APPENDIX A. GPT PROMPT FOR GENERATING PARAPHRASES IN THE AUGMENTED POST-EDITING CONDITION}

\noindent Generate two paraphrases for this message by following the instructions:

\noindent The message is about two houses that the person is considering purchasing;

\noindent Use natural English expressions and sentence structure in your paraphrases;

\noindent Adjust the message’s wording to make one paraphrase more casual and one more formal;

\noindent The length of the paraphrase should be within a 5-word difference from the original message.

\end{document}